\documentclass[useAMS,usenatbib]{mn2e}
\usepackage{natbib}
\usepackage{graphicx}
\newcommand{\nat}{Nature}
\newcommand{\mnras}{MNRAS}
\newcommand{\apj}{ApJ}
\newcommand{\apjl}{ApJL}
\newcommand{\apjs}{ApJS}
\newcommand{\aap}{A\&A}

\newcommand{\prd}{Phys.Rev.D}
\def \ssr #1 #2 {{ Space\ Sci.\ Rev.\/} {\bf #1} #2}
\newcommand{\eqb}{\begin{equation}}
\newcommand{\eqe}{\end{equation}}
\newcommand{\sgn}{\mathop{\mathrm{sgn}}}
\begin{document}

\title{Radio emission of the Crab and Crab-like pulsars}
\author[Yuri Lyubarsky]{Yuri Lyubarsky\\
Physics Department, Ben-Gurion University, P.O.B. 653, Beer-Sheva
84105, Israel; e-mail: lyub@bgu.ac.il}
\date{Received/Accepted}
\maketitle
\begin{abstract}
 The pulsar radio emission  is commonly associated with
the plasma outflow in the open field line tube; then a pencil beam is emitted along the pulsar
magnetic axis. Observations suggest that there is an additional radio emission mechanism specific
for pulsars with high magnetic field at the light cylinder. These pulsars are known to be strong
sources of non-thermal high energy radiation, which could be attributed to reconnection in the
current sheet separating, just beyond the light cylinder, the oppositely directed magnetic fields.
Pulsars with the highest magnetic field at the light cylinder ($>100$ kG)
exhibit also radio pulses in phase with the high energy pulses. Moreover, giant radio
pulses are observed in these pulsars. 
I argue that the reconnection process  that produces high energy emission could also be responsible
for the radio emission. Namely, coalescence of magnetic islands in the sheet produces magnetic
perturbations that propagate away in the form of electro-magnetic nano-shots. I estimate the
parameters of this emission and show that they are compatible with observations.
\end{abstract}
\begin{keywords}
 magnetic reconnection -- (magnetohydrodynamics) MHD -- plasmas -- radiation mechanisms: non-thermal --
(stars:) pulsars: general
\end{keywords}

\section{Introduction}

Pulsars with large magnetic fields at the light cylinder differ significantly from ordinary
pulsars. First of all,   non-thermal high-energy emission is observed only from pulsars with the
field at the light cylinder of about or larger than 1 kG (see, e.g., fig.\ 2 in the review by
\citealt{Venter18}). Moreover, the radio properties of many of them differ significantly from the
properties of ordinary pulsars. The Crab pulsar is a good example. It exhibits two well separated
main peaks of radio emission.
 Taking into account that the pulsar rotation axis is inclined to the line of sight by $60^o$, as
follows from the X-ray image of the nebula (e.g., \citealt{Ng_Romani04}), this is incompatible with
the standard presumption that the pulsar radio emission forms a pencil beam generated in the polar
cap outflow and therefore directed along the magnetic axis. In the standard picture, two-peaked
profiles could be formed either in orthogonal
 or in aligned rotators (in the last case, the double peak is formed by a hollow cone beam).
 Moreover, if pulses of ordinary pulsars generally widen
with decreasing frequency (e.g., \citealt{Graham-Smith03}), the Crab main pulses do not exhibit any
significant frequency evolution. One more specific feature of the Crab radio emission is giant
pulses.  The Crab twin in the LMC, PSR B0540-69, exhibits  similar properties (e.g.,
\citealt{Johnston_etal04}).
All this evidences for an emission mechanism different from that responsible for the "standard"
pulsar emission.

It seems that the same specific emission mechanism operates also in many recycled (millisecond)
pulsars because properties of their radio emission \citep{Kramer_etal98} resemble those of the
Crab. First of all, the abundance of double pulses in recycled pulsars is incompatible with the
standard explanation of double-peaked pulsars as orthogonal rotators or nearly aligned rotators
with a hollow cone beam. Moreover, there is no or very small frequency development of the pulses.
In millisecond pulsars, the beam width is less than what is predicted by the canonical pulsar model
and in some cases is even less than the size of the polar cap.  The millisecond pulsars are also
known as gamma-ray sources and in some of them, giant pulses are observed. One sees a similarity
between the Crab-like and the recycled pulsars; therefore there should be a special radio emission
mechanism associated with the high-energy emission region of pulsars. Of course the "standard"
radio emission mechanism could also operate in these pulsars so that, depending on the orientation,
the observer sees radiation either from the polar tube outflow, or from the site where
the high energy emission is generated, 
or from both. For example in the Crab pulsar, the "standard" mechanism may be responsible for the
precursor radio emission (e.g., \citealt{Graham-Smith03}).

The common feature of all these pulsars is a large magnetic field at the light cylinder. The model
successfully explaining the high energy emission of pulsars places the emission source to the
current sheet separating, just beyond the light cylinder, the oppositely directed magnetic fields
\citep{Lyubarsky96,Bai_Spitkovsky09,Arka_Dubus12,Kalapotharakos_etal12,Kalapotharakos_etal17,Cerutti_etal16}.
The magnetic reconnection heats the particles in the sheet and if the magnetic field in the light
cylinder zone is high enough, the synchrotron emission of the particles forms a powerful
high-energy fan beam. Inasmuch as such a beam rotates together with the magnetosphere, the observer
generally sees a double-peaked light curve.

In this paper, I propose a mechanism for the radio emission from a reconnecting current sheet in
pulsars. The reconnection process occurs via formation and coalescence of magnetic islands/pinches
within the current sheet (e.g., \citealt{Kagan_etal15}). Two islands coalescing with the velocity
of the order of the speed of light perturb the magnetic field in the vicinity of the coalescence
point thus producing magneto-hydrodynamic (MHD) waves propagating away. I show that fast
magneto-sonic (fms) waves escape from the magnetosphere in the form of radio waves.  In particular,
the coalescence of large islands could be responsible for giant nano-shots observed in the Crab
pulsar \citep{Hankins_etal03,Eilek_Hankins16}.

The paper is organized as follows. In sect. 2, I present the emission mechanism and estimate the
properties of nano-shots produced by a coalescence event. In sect. 3, I consider non-linear
interactions between waves generated by many coalescence events and estimate the radio luminosity
provided by the proposed mechanism.
Discussion and conclusions are presented in sect. 4.

\section{Radio emission from  magnetic islands coalescing in the pulsar current sheet}

Coalescence of two magnetic islands produces a magnetic perturbation in the vicinity of the
coalescence region. Therefore MHD waves are generated around the reconnecting current sheet. There
are generally three MHD waves, the Alfven wave and two  magnetosonic waves. In the relativistic
case, their phase velocities are presented, e.g., by \citet{Appl_Camenzing88}. In the simplest case
of a cold plasma, only the Alfven and the fast magnetosonic (fms) waves remain; their phase
velocities are reduced to
\begin{eqnarray}
v_A=c\sqrt{\frac{\sigma}{1+\sigma}}\,\cos\theta;\\
 v_{\rm fms}=c\frac{\sigma}{\sqrt{1+\sigma^2}}; \label{fms-disp}
\end{eqnarray}
where $\theta$ is the angle between the magnetic field and the direction of propagation,
$\sigma=B^2/4\pi\rho c^2$ the magnetization parameter, $B$ the background magnetic field, $\rho$
the plasma density. The group velocity of the Alfven waves is directed along the magnetic field
lines therefore, they do not transfer the energy away from the current sheet. Fast magnetosonic
waves do propagate across the magnetic field lines therefore any coalescence event produces a
quasi-spherical fms pulse of the duration  $\sim a/c$, where $a$ is the transverse size of the island.

 In order to demonstrate that this wave escapes from the system in the form of a vacuum
electromagnetic wave, let us consider what happens to this wave when it propagates towards smaller
plasma densities.   In the harmonic wave with the frequency $\omega$ and the wave vector
$\mathbf{k}$, the electric current is found from Maxwell's equations as \eqb
\mathbf{j}=\frac{i}{4\pi}\left(\frac{\mathbf{k(k\cdot
E)}-k^2\mathbf{E}}{\omega}+\omega\mathbf{E}\right),
\eqe
where $\mathbf{E}$ is the electric field of the wave. In the fms wave, the electric field is
perpendicular both to the background magnetic field and to the direction of propagation therefore
$\mathbf{k\cdot E}=0$. 
Then the ratio of the conductivity to the displacement current in the fms wave is presented, with
the aid of eq. (\ref{fms-disp}), as
   \eqb
 \frac{j}{i\omega E}=\frac 1{4\pi (1+\sigma^2)}.
   \eqe
One sees that  when the ratio of the plasma to the magnetic energy density goes to zero,
$\sigma\to\infty$, the conductivity current vanishes. 
Therefore the wave smoothly transforms to a vacuum
electro-magnetic wave when the plasma density goes to zero.

More generally, one can abandon the MHD approximation and consider the wave in the scope of the
two-fluid hydrodynamics. For the electron-positron plasma,  the dispersion relation for
 waves  polarized perpendicularly both to the background
magnetic field and to the direction of propagation is found as  (e.g., \citealt{Melrose97})
 \eqb
 \omega^2=k^2c^2-\frac{2\omega_p^2\omega^2}{\omega_B^2-\omega^2},
 \eqe
where $\omega_p=\sqrt{4\pi e^2n/m}$ is the plasma frequency, $\omega_B=eB/mc$ the Larmor frequency,
$n$ the electron density, $e$ and $m$ the electron charge and mass, correspondingly. Taking into
account that $\sigma=\omega^2/(2\omega_p^2)$, one sees that this equation is reduced to eq.
(\ref{fms-disp}) in the limit $\omega,\,\omega_p\ll\omega_B$.

Numerical simulations of pulsar magnetospheres \citep{Philippov_etal15a} show that the plasma fills
the magnetosphere and the wind inhomogeneously such that empty regions remain that propagate
outwards. When the fms wave enters these regions, it becomes truly vacuum electro-magnetic wave. If
the wave remains within the plasma, it eventually meets the cyclotron resonance where it could be
absorbed, at least partially.  The optical depth for the cyclotron absorption depends on the plasma
parameters in this region
\citep{Blandford_Scharlemann76,Mikhailovskii_etal82,Lyubarsky_Petrova98,Luo_Melrose01,Petrova02,Fussel_etal03},
 which lies well outside the light cylinder in pulsars with a high magnetic field at the light cylinder.
 The question of how the radio emission
passes through the resonance region is common for all pulsar radiation models; it is not addressed
here. For our purpose, it is enough to notice that even if the radio emission is effectively
absorbed by the plasma in the resonance layer, some radiation still could escape due to the
mentioned above empty regions.

Let us now consider parameters of the emitted pulses. The pulse width in the frame moving with the
plasma within the current sheet is of the order of the transverse size of magnetic islands, $a$. In
the lab frame, the duration of the pulse is
 \eqb
 \tau\sim \frac a{c\Gamma},
 \label{duration}\eqe
where $\Gamma$ is the Lorentz factor of the plasma  flow within the sheet.  According to the
available  models (e.g., \citealt{Timokhin_Arons13}), the magnetospheric plasma moves outwards
along the magnetic field lines with Lorentz factors $\sim 100 - 1000$. When the plasma enters, via
the reconnection process, into the current sheet, it is decelerated because the magnetic field
lines now cross the sheet giving rise to the decelerating $\mathbf{j\times B}$ force.  Therefore
one can expect that $\Gamma$ is less than the Lorentz factor of the plasma flow in the
magnetosphere. It was found \citep{Lyubarsky96} that at $\Gamma\sim 10$, parameters of the high
energy emission from the current sheet are compatible with the observed parameters. Therefore
$\Gamma$ will be normalized as $\Gamma=10\Gamma_1$.

The size of magnetic islands scales with the width of the current sheet, $\Delta$. In pulsars with
high magnetic fields at the light cylinder, the width of the sheet is determined by the balance
between the dissipative heating of the plasma in the sheet
and the synchrotron cooling. According to the available rough estimates
\citep{Lyubarsky96,Uzdensky_Spitkovsky14},
 \eqb
 \Delta\sim r_e^{-1/2}\left(\frac c{\omega_B}\right)^{3/2}= 1.3 B^{-3/2}_6\rm m,
 \label{sheet-width}\eqe
where $r_e$ is the classical electron radius, $B=10^6B_6$ G the magnetic field at the light
cylinder.

Numerical simulations of the reconnection process show \citep{Petropoulu16,Petropoulu18} that the
size of islands spans a wide range. The large islands are ten or even more times larger than the
width of the sheet therefore, $a$ will be normalized by 10 m, $a=10^3a_3$ cm. Now the observed
duration of the pulse, eq. (\ref{duration}), is estimated as
 \eqb
 \tau\sim 3\frac{a_3}{\Gamma_1}\rm ns.
 \eqe
The corresponding frequency is $f\sim \tau^{-1}\sim 3\cdot 10^8$ Hz. In principle, one can expect
that after the coalescence of two large islands, the newly born island oscillates therefore one can
write, more generally, $f\tau\sim$ few. Recall that the observed nano-shots exhibit $f\tau\sim 10$
\citep{Hankins_etal03,Eilek_Hankins16} however, such fine details could not be captured by the
presented rough model.

The amplitude of the magnetic perturbation produced by the coalescence of two large  islands is
comparable with the strength of the background field therefore the total energy of the pulse may be
estimated as the energy density of the background field multiplied by the volume of the island.
Taking into account that the islands are in fact current ropes elongated in the direction of the
current, one can take the length of the rope, $l=\zeta a$, where $\zeta>1$; then the energy of the
pulse in the frame moving with the plasma in the current sheet is
 \eqb
 {\cal E'}\sim \frac{B^2}{8\pi}la^2=\frac{\zeta B^2}{8\pi}a^3.
 \eqe
In the lab frame, ${\cal E=E'}\Gamma$. This energy is emitted within the solid angle $\sim
\pi\Gamma^{-2}$ during the time $\tau$. The spectral flux detected at the distance $D=2D_2$ kpc is
 \eqb
 S\sim\frac{{\cal E}\Gamma^2}{\pi f\tau D^2}\sim\frac{\zeta\Gamma^3B^2a^3}{8\pi^2D^2}=
 350\frac{\zeta_1\Gamma^3_1B_6^2a_3^3}{D^2_{2}}\, \rm Jy,
 \eqe
where $\zeta=10\zeta_1$. One sees that the estimated duration and flux are compatible with the
observed parameters of giant nano-shots from the Crab pulsar
\citep{Hankins_etal03,Eilek_Hankins16}.

The above estimates assume that the pulses freely escape and reach the observer. However, pulses
from different coalescence events generally intersect each other above the current sheet. This
would lead to non-linear interaction, which is considered in the next section.

\section{Non-linear interactions of fms waves}
An fms pulse produced by a coalescence events could pass through pulses produced in coalescence
events throughout the sheet. Therefore the  non-linear interaction between the pulses should be
generally taken into account. The simplest is the interaction of three waves (e.g.,
\citealt{Tsytovich70}) satisfying the resonance conditions
\eqb
\omega_1+\omega_2=\omega;\qquad \mathbf{k}_1+\mathbf{k}_2=\mathbf{k}.
\label{resonance}\eqe
Let us consider this process in the force-free limit because outside the currents sheet, the
magnetic field energy significantly exceeds the plasma energy density.

In this case, the fms and the Alfven waves have the dispersion relations (see, e.g., Appendix)
\eqb
\omega=kc
\label{fms}\eqe
and
\eqb
\omega=kc\vert\cos\theta\vert, \label{Alfven}\eqe
 correspondingly, where $\theta$ is the angle between the
wave vector and the background magnetic field. Three fms waves could not satisfy the resonance
conditions (\ref{resonance})\footnote{More exactly, three fms waves could satisfy the conditions
(\ref{resonance}), if they are aligned. But it is shown in Appendix, that even in this case, the
non-linear interaction vanishes because in the force-free case, fms waves do not excite either
currents or charge in the plasma.} however, an fms wave could decay
into another fms wave and an Alfven wave or into two Alfven waves.

When considering non-linear processes, one can conveniently use the wave amplitudes,
$a_{\mathbf{k}}$, defined such that
\eqb
n_{\mathbf{k}}=\vert a_{\mathbf{k}}\vert^2
 \label{n}\eqe
 is the number density of quanta with the wave vector $\mathbf{k}$, the wave energy density being
\eqb
{\cal E}_{\mathbf{k}}=\omega_{\mathbf{k}} n_{\mathbf{k}}. \label{n} \eqe The electric and magnetic
fields are equal in the force-free fms and Alfven waves (see, e.g., Appendix). Therefore for a
harmonic wave,
 \eqb
 \mathbf{E}=\mathbf{E}_{\mathbf{k}}\exp(i\mathbf{k\cdot r}-i\omega_{\mathbf{k}} t)+\rm c.c.,
 \label{harmonics}\eqe
 the average wave energy density is just
 \eqb
 {\cal E}_{\mathbf{k}}=\frac{\overline{\mathbf{E}^2}}{4\pi}=\frac{\vert \mathbf{E}_{\mathbf{k}}\vert^2}{2\pi}.
 \label{energy}\eqe
 Then
\eqb
a_{\mathbf{k}}=\frac{E_{\mathbf{k}}}{\sqrt{2\pi\omega_{\mathbf{k}}}}.
\label{a}\eqe

Due to the three-wave interaction, the amplitudes of waves satisfying the resonance condition
(\ref{resonance}) vary according to equations
\begin{eqnarray}
\frac{\partial a_{\mathbf{k}_1}}{\partial t}=V_1 a^*_{\mathbf{k}_2}a_{\mathbf{k}};\label{adot1}\\
\frac{\partial a_{\mathbf{k}_2}}{\partial t}=V_2 a^*_{\mathbf{k}_1}a_{\mathbf{k}};\label{adot2}\\
\frac{\partial a_{\mathbf{k}}}{\partial t}=V_3 a_{\mathbf{k}_1}a_{\mathbf{k}_2}\label{adot3}.
\end{eqnarray}
In the three-wave process, annihilation of a quantum $\mathbf{k}$ results in creation of one
quantum $\mathbf{k}_1$ and one quantum $\mathbf{k}_2$, so that $n_{\mathbf{k}}+n_{\mathbf{k}_1}=\it
const$; $n_{\mathbf{k}}+n_{\mathbf{k}_2}=\it const$ (the Manley-Rowe relations); this implies
\eqb
V_3=-V^*_2=-V^*_1\equiv V_{\mathbf{k}_1\mathbf{k}_2\mathbf{k}}.
\eqe
The matrix elements, $V_{\mathbf{k}_1\mathbf{k}_2\mathbf{k}}$, for the interaction of an fms wave
with two fms waves or with  an Alfven wave and an fms wave  are calculated in Appendix.

The decay time of a monochromatic fms wave $(\omega, \mathbf{k})$ could be estimated assuming that
initially $a_{\mathbf{k}_1}$, $a_{\mathbf{k}_2}\ll a_{\mathbf{k}}$. Then eliminating, e.g.,
$a_{\mathbf{k}_2}$ from eqs. (\ref{adot1}) and (\ref{adot2}), one gets
\eqb
\frac{\partial^2a_{\mathbf{k}_1}}{\partial t^2}=
\vert V_{\mathbf{k}_1\mathbf{k}_2\mathbf{k}}a_{\mathbf{k}}\vert^2a_{\mathbf{k}_1}.
\eqe
One sees that $a_{\mathbf{k}_1}$ (and also $a_{\mathbf{k}_2}$) grow exponentially with the
characteristic time
\eqb
\tau=\left(\vert V_{\mathbf{k}_1\mathbf{k}_2\mathbf{k}_3}a_{\mathbf{k}_3}\vert^2\right)^{-1/2}.
\eqe
This means that at the time scale of the order of a few $\tau$, the waves $\mathbf{k}_1$ and
$\mathbf{k}_2$ take a significant fraction of the initial energy. Therefore the amplitude of the
initial wave, $a_{\mathbf{k}}$, decreases at the same time-scale. Substituting the estimate
(\ref{V}) for the matrix element, one gets an estimate for the decay time
\eqb
\tau\sim \left(\frac{\vert E_{\mathbf{k}}\vert}{B_0}\omega\right)^{-1}.
\eqe

In eqs. (\ref{adot1}-\ref{adot3}), the waves are assumed to be monochromatic with fixed phases. In
a more realistic case of wide spectra and random phases, one presents the electric field as a
superposition of harmonics,
 \eqb
 \mathbf{E}=\int\left[\mathbf{E}_{\mathbf{k}}\exp(i\mathbf{k\cdot r}-i\omega t)+{\rm c.c.}\right]d\mathbf{k},
 \eqe
such that the wave energy density is
\eqb
{\cal E}=\int {\cal E}_{\mathbf{k}}d\mathbf{k}, \eqe
 where the energy of each harmonic, ${\cal E}_{\mathbf{k}}$, is given by eq. (\ref{energy}).

In this case, the wave field is described by the number density of quanta defined by eq. (\ref{n}),
the evolution being governed by the kinetic equations
 \begin{eqnarray}
\frac{\partial n_{\mathbf{k}}}{\partial t}=\int W_{\mathbf{k}_1\mathbf{k}_2\mathbf{k}}(n_{\mathbf{k}_1}n_{\mathbf{k}_2}-
n_{\mathbf{k}_1}n_{\mathbf{k}}\nonumber\\
-n_{\mathbf{k}_2}n_{\mathbf{k}})
\delta(\omega_{\mathbf{k}}-\omega_{\mathbf{k}_1}-\omega_{\mathbf{k}_2})d^3\mathbf{k}_2,
 \end{eqnarray}
 where
 \eqb
 W_{\mathbf{k}_1\mathbf{k}_2\mathbf{k}}=2\pi\vert V_{\mathbf{k}_1\mathbf{k}_2\mathbf{k}}\vert^2.
 \eqe
The characteristic interaction time may be now estimated as
\eqb
\tau\sim \left(W_{\mathbf{k}_1\mathbf{k}_2\mathbf{k}} n_{\mathbf{k}}k^2\right)^{-1}\sim\left(\frac{\cal E}{U_0}\omega\right)^{-1},
\label{decay_time}\eqe
where $U_0=B_0^2/8\pi$ the energy density of the background field; ${\cal E}=\omega_{\mathbf{k}}n_{\mathbf{k}}k^3$ the energy density of the waves.

One sees that the interaction of fms waves inevitably produces Alfven waves.
The Alfven waves evolve into a cascade transferring the energy to small scales where they
eventually decay and heat the plasma \citep{Thompson_Blaes98}. Therefore if the time scale
(\ref{decay_time}) is smaller than the characteristic time of the system (e.g., the escape time
from the system), most of the initial wave energy is eventually converted to heat.

If the reconnection in the current sheet just beyond the light cylinder proceeds continuously
generating many fms pulses from local coalescing events, the interaction between the pulses would
transfer their energy to the plasma until the characteristic wave interaction time
(\ref{decay_time}) becomes equal to the escape time,
$\tau\sim R_L/c=\Omega^{-1}$. Then the radiation energy density is ${\cal E}\sim (\Omega
/\omega)U_0$.
Therefore  the total luminosity is estimated\footnote{The non-linear interactions in the force-free
regime are estimated in the zero electric field frame; it is not affected by plasma moving along
the magnetic field lines. Just beyond the light cylinder, the magnetospheric electric and magnetic
fields are of the same order but not too close to each other so that the drift velocity, $cE/B$, is
only mildly relativistic. Therefore to within factors of the order of unity, the parameters in the
zero electric field frame and in the lab frame are the same.}. as
\eqb
L\sim {\cal E}cR^2_L\sim \frac{ U_0R_Lc^2}{\omega}.
\eqe
Taking into account that the pulsar spin-down power is estimated as $L_{\rm sd}\sim U_0cR_L^2$, one
can present the luminosity in the form
\eqb
L\sim\frac{\Omega}{\omega}L_{\rm sd}=\frac{L_{\rm sd}}{ Pf},
\eqe
where $P=2\pi/\Omega$ is the pulsar rotation period, $f$ the radio frequency. Substituting the Crab
rotational period and $f=100$ MHz (the Crab pulsar spectrum is very steep so that most of the
energy is emitted at low frequencies), one gets $L/L_{\rm sd}\sim 3\cdot 10^{-7}$, which is roughly
compatible with the observed Crab radio luminosity
$L\approx 7\cdot  10^{31}$ erg/s (e.g., \citealt{Malov_etal94}).

\section{Discussion and conclusions}

In this paper, I consider the radio emission generated by coalescence of magnetic islands in a
reconnecting pulsar current sheet just beyond the light cylinder. A coalescence event produces a
short fms pulse that is smoothly converted into an electro-magnetic wave when propagates towards
the decreasing plasma density. The duration of the pulses, and therefore the effective emission
wavelength, depends on the size of the islands, which scale with the width of the current sheet.
The last is determined by the balance between the dissipative heating and synchrotron cooling.
According to the estimate (\ref{sheet-width}), the width of the sheet rapidly grows with decreasing
of the magnetic field strength, therefore the proposed mechanism works only in pulsars with a high
enough magnetic field at the light cylinder.

Already for the Vela pulsar parameters, one gets the sheet width of about 100 m.
Therefore most of the emission is expected to be in an extremely low frequency band. Taking into
account that islands of smaller sizes are presented in the current sheet, one cannot exclude that
in this case, the emission from the current sheet still could be observed in the decameter range.
But this emission could hardly be observed from pulsars with larger periods.
Similar considerations show that this emission could be observed from  millisecond pulsars with the
highest values of the magnetic field at the light cylinder. This is indeed the case. In all pulsars
that exhibit radio and gamma peaks aligned, which evidences for the radio emission from the current
sheet, the estimated magnetic field at the light cylinder exceeds $100$ kG
\citep{Johnson_etal14,Ng_etal14}. The giant pulses are also observed only from millisecond pulsars
with $B>100$ kG \citep{Bilous_etal15}. Therefore the presented model is consistent with
observations.

It was shown in this paper that at the Crab pulsar parameters, the coalescence of large magnetic
islands produces nano-shots with the energy and duration compatible with the observed giant pulses.
When many nano-shots are continuously produced in the current sheet, non-linear interactions
between them transform most of the energy into heat, the average luminosity being determined by the
condition that the system (in our case, the near zone of the pulsar wind with the size of the order
of the light cylinder radius) is marginally transparent for non-linear interactions. The Crab radio
luminosity estimated from these considerations is compatible with that observed.


\section*{Acknowledgments}
The work was supported by the grant I-1362-303.7/2016 from the German-Israeli Foundation for
Scientific Research and Development.


\appendix
\section*{Appendix. Three-wave interactions in the force-free MHD}
\setcounter{equation}{0}
\renewcommand\theequation{A.\arabic{equation}}
The general theory of the nonlinear wave interactions in the force-free MHD has been developed by
\citet{Thompson_Blaes98}. Here I calculate straightforwardly the interaction rates.

The force-free MHD equations are written as
\begin{eqnarray}
\rho\mathbf{E}+\frac 1c\mathbf{j\times B}=0;\label{force}\\
\mathbf{E\cdot B}=0.
\label{MHD}
\end{eqnarray}
They should be complemented by Maxwell's equations
\begin{eqnarray}
\nabla\cdot\mathbf{E}=4\pi\rho;\qquad \nabla\times\mathbf{E}=-\frac 1c\frac{\partial \mathbf{B}}{\partial t};\\
\nabla\cdot\mathbf{B}=0;\quad\nabla\times\mathbf{B}=\frac{4\pi}c\mathbf{j}+\frac 1c\frac{\partial \mathbf{E}}{\partial t}.
\end{eqnarray}
Let the background magnetic field be directed along $z$-axis, $\mathbf{B}_0=B_0\hat{\mathbf{z}}$.
Assuming that perturbations are small, one can solve eqs. (\ref{force}) and (\ref{MHD})
perturbatively, $\mathbf{E}=\mathbf{E}^{(1)}+\mathbf{E}^{(2)}+\dots$;
$\mathbf{B}=B_0\hat{\mathbf{z}}+\mathbf{B}^{(1)}+\mathbf{B}^{(2)}+\dots$.

Linearizing eqs. (\ref{force}) and (\ref{MHD}) yields
\begin{eqnarray}
 \mathbf{j}^{(1)}\times\hat{\mathbf{z}}=0;\label{force1}\\
E_z^{(1)}=0.\label{Ez=0}
\end{eqnarray}
Making use of Maxwell's equations, one expresses the current, $\mathbf{j}^{(1)}$,  via the fields;
then eq. (\ref{force1}) is written, for a harmonic wave, as
\eqb
[(\omega^2-k^2c^2)\mathbf{E}^{(1)}_{\mathbf{k}}+c^2\mathbf{(k\cdot
E}^{(1)}_{\mathbf{k}})\mathbf{k}]\times\hat{\mathbf{z}}=0. \label{jxz=0}\eqe The set of equations
(\ref{Ez=0}) and (\ref{jxz=0}) has two solutions.

The first solution is polarized perpendicularly to the $\mathbf{kB}_0$ plane so that one can write
the electric field of the wave as
\eqb
\mathbf{E}^{(1)}_{\mathbf{k}}=\frac{\hat{\mathbf{z}}\times\mathbf{k}}{k\sin\theta}E_{\mathbf{k}}^{(1)};
\label{Efms}\eqe
 where $\theta$ is the angle between the wave vector and the background magnetic
field. Then eq. (\ref{jxz=0}) yields
\eqb
\omega=kc. \label{fms2}\eqe
 Substituting this solution into Maxwell's equations, one gets
 \begin{eqnarray}
 \mathbf{B}^{(1)}_{\mathbf{k}}=\frac 1k\mathbf{k}\times \mathbf{E}^{(1)}_{\mathbf{k}}
 =\frac{\mathbf{\hat{z}}-\cos\theta\mathbf{k}}{k\sin\theta}E^{(1)}_{\mathbf{k}};\label{Bfms}\\
 \quad \rho^{(1)}=\mathbf{j}^{(1)}=0. \label{fms1}
 \end{eqnarray}
  This is the fms wave.

The second solution is polarized in the $\mathbf{kB}_0$ plane, so that one writes
\eqb
\mathbf{E}^{(1)}_{\mathbf{k}}=\frac{\mathbf{k}-\hat{\mathbf{z}}k\cos\theta}{k\sin\theta}E_{\mathbf{k}}^{(1)}\label{EAlfven}.
\eqe
In this case,
 \begin{eqnarray} \omega=kc\vert\cos\theta\vert;\label{Alfven2}\\
 \mathbf{B}^{(1)}_{\mathbf{k}}=\frac{(\mathbf{k\cdot\hat{z}})}{\omega
 k\sin\theta}\mathbf{\hat{z}\times k}\, E^{(1)}_{\mathbf{k}};\\
\rho_{\mathbf{k}}^{(1)}=\frac{ik\sin\theta E_{\mathbf{k}}^{(1)}}{4\pi};\quad
\mathbf{j}_{\mathbf{k}}^{(1)}=\sgn(\cos\theta)c\rho_{\mathbf{k}}^{(1)}\hat{\mathbf{z}};.\label{Alfven1}
\end{eqnarray}
This is the Alfven wave.

In the second order, eqs. (\ref{force}) and (\ref{MHD}) are written as
\begin{eqnarray}
\rho^{(1)}\mathbf{E}^{(1)}+\frac 1c\mathbf{j}^{(1)}\times\mathbf{B}^{(1)}+
\mathbf{j}^{(2)}\times\hat{\mathbf{z}}B_0=0\label{force2}\\
\mathbf{E}^{(1)}\cdot\mathbf{B}^{(1)}+B_0\mathbf{E}^{(2)}\hat{\mathbf{z}}=0.\label{MHD2}
\end{eqnarray}
In this order, the wave amplitudes slowly vary with time therefore Maxwell's equations should be
presented, in Fourier components, as
\begin{eqnarray}
i\omega\mathbf{B}^{(2)}_{\mathbf{k}}-\frac{\partial\mathbf{B}^{(1)}_{\mathbf{k}}}{\partial t}=i\mathbf{k\times E}^{(2)}_{\mathbf{k}};\\
\frac{4\pi}c\mathbf{j}^{(2)}_{\mathbf{k}}=i\mathbf{k\times B}^{(2)}_{\mathbf{k}}+i\omega\mathbf{E}^{(2)}_{\mathbf{k}}-
\frac{\partial\mathbf{E}^{(1)}_{\mathbf{k}}}{\partial t}\\
=\frac i{\omega}\left\{\right(\omega^2-k^2)\mathbf{E}^{(2)}_{\mathbf{k}}+(\mathbf{k\cdot E}^{(2)}_{\mathbf{k}})\mathbf{k}\}\nonumber\\
+\frac 1{\omega^2}
\left\{\left(\mathbf{k}\cdot \frac{\partial\mathbf{E}^{(1)}_{\mathbf{k}}}{\partial t}\right)\mathbf{k}-(\omega^2+k^2)\frac{\partial\mathbf{E}^{(1)}_{\mathbf{k}}}{\partial t}\right\}.
\label{j2}\end{eqnarray}

The Fourier transform of eq. (\ref{force2}) yields
\eqb
\sum_{\mathbf{k'}}\left(\rho^{(1)}_{\mathbf{k'}}\mathbf{E}^{(1)}_{\mathbf{k-k'}}+
\mathbf{j}^{(1)}_{\mathbf{k'}}\times\mathbf{B}^{(1)}_{\mathbf{k-k'}}\right)+
B_0\mathbf{j}^{(2)}_\mathbf{k}\times\hat{\mathbf{z}}=0.
\eqe
For waves satisfying the resonance condition (\ref{resonance}), it is reduced to
\eqb
\rho^{(1)}_{\mathbf{k}_1}\mathbf{E}^{(1)}_{\mathbf{k}_2}+
\rho^{(1)}_{\mathbf{k}_2}\mathbf{E}^{(1)}_{\mathbf{k}_1}+
\mathbf{j}^{(1)}_{\mathbf{k}_1}\times\mathbf{B}^{(1)}_{\mathbf{k}_2}+
\mathbf{j}^{(1)}_{\mathbf{k}_2}\times\mathbf{B}^{(1)}_{\mathbf{k}_1}=
B_0\hat{\mathbf{z}}\times\mathbf{j}^{(2)}_\mathbf{k}.
 \label{2order}\eqe
 Now let us consider specific cases.

{\bf Interaction of three fms waves.} In fms waves, the current and charge density vanish in the
first approximation, therefore for three fms waves, eq. (\ref{force2}) is reduced to
\eqb
\mathbf{j}^{(2)}\times\hat{\mathbf{z}}=0.
 \label{3fms}\eqe
 The non-linear current (\ref{j2}) for
the fms wave is found, by applying eqs. (\ref{Efms}) and (\ref{fms2}), as
\eqb
\frac{4\pi}c\mathbf{j}^{(2)}_{\mathbf{k}}=i\frac{(\mathbf{k\cdot
E}^{(2)}_{\mathbf{k}})\mathbf{k}}k-\frac{2}{k\sin\theta}\frac{\partial
E^{(1)}_{\mathbf{k}}}{\partial t}\mathbf{\hat{z}\times k}.
 \label{j2fms}\eqe
Substituting this expression into eq. (\ref{3fms}) and making a dot product of this equation with
$\mathbf{k}$ in order to kill the term with $\mathbf{E}^{(2)}_{\mathbf{k}}$ yields $\frac{\partial
E^{1}_{\mathbf{k}}}{\partial t}=0$ therefore three fms waves do not interact.

{\bf Decay of an fms wave into an fms and an Alfven waves.} Taking into account eqs. (\ref{fms1})
and (\ref{Alfven1}), one writes eq. (\ref{2order}) as
\eqb
\rho^{(1)}_{\mathbf{k}_1}\left(\mathbf{E}_{\mathbf{k}_2}^{(1)}+
\sgn(\cos\theta_1)\hat{\mathbf{z}}\times
\mathbf{B}_{\mathbf{k}_2}^{(1)}\right)=B_0\hat{\mathbf{z}}\times\mathbf{j}^{(2)}_{\mathbf{k}};
\eqe
where $\mathbf{k}_1$ is for the Alfven wave and $\mathbf{k}_2$ for the fms wave. Making a dot
product of this equation with $\mathbf{k}$
and applying eqs. (\ref{Efms}-\ref{Bfms}), (\ref{EAlfven}-\ref{Alfven1}) and (\ref{j2fms}) yields
\begin{eqnarray}
\frac{\partial E^{(1)}_{\mathbf{k}}}{\partial t} =
-i\frac{ck_1\sin\theta_1[1-\sgn(\cos\theta_1)\cos\theta_2]}
{2B_0k_2k\sin\theta_2\sin\theta}\\
\times [(\mathbf{k}_2\times\mathbf{k})\cdot\hat{\mathbf{z}}]
E^{(1)}_{\mathbf{k}_1}E^{(1)}_{\mathbf{k}_2}.
\end{eqnarray}
This equation is reduced to the form of eq. (\ref{adot3}) by transforming, according to eq. (\ref{a}), the wave amplitudes  from $E_\mathbf{k}$ to $a_\mathbf{k}$; then one gets the expression for the matrix coefficient
\begin{eqnarray}
V^{\rm S\to S+A}_{\mathbf{k}_1\mathbf{k}_2\mathbf{k}}=
-i\left(\frac{\pi\vert\cos\theta_1\vert}{2k_2}\right)^{1/2}\left(\frac{k_1c}{k}\right)^{3/2}\label{V}\\
\times\frac{\sin\theta_1[1-\sgn(\cos\theta_1)\cos\theta_2][(\mathbf{k}_2\times\mathbf{k})\cdot\hat{\mathbf{z}}]}
{B_0\sin\theta_2\sin\theta}.\nonumber
\end{eqnarray}

{\bf Decay of an fms wave into two Alfven waves.} Now eq. (\ref{2order}) is written, with account
of eq. (\ref{Alfven1}), as
 \begin{eqnarray}
\left\{\rho^{(1)}_{\mathbf{k}_1}\mathbf{E}_{\mathbf{k}_2}^{(1)}+
\rho^{(1)}_{\mathbf{k}_2}\mathbf{E}_{\mathbf{k}_1}^{(1)}\right\}
\left[1-\sgn(\cos\theta_1)\sgn(\cos\theta_2)\right]\nonumber\\
=B_0\hat{\mathbf{z}}\times\mathbf{j}^{(2)}_{\mathbf{k}},
 \label{AtoA+A}\end{eqnarray}
 where both $\mathbf{k}_1$  and $\mathbf{k}_2$ are for the Alfven waves.
 One sees that the decay is possible only if the Alfven waves propagate in the opposite directions
 with respect to the background magnetic field,
 \eqb
\cos\theta_1\cos\theta_2<0.
 \eqe
 This is a partial case of the general fact that Alfven waves propagating in the same direction do
 not interact with each other.
Let us assume that this condition is satisfied; then making a dot product of eq. (\ref{AtoA+A})
with $\mathbf{k}$ and applying eqs.
 (\ref{Alfven2}-\ref{Alfven1}) and (\ref{j2fms}), one gets
 \eqb
\frac{\partial E^{(1)}_{\mathbf{k}}}{\partial t} =
i\frac{c}{B_0}\left(k_1\sin\theta_1\cos\phi_2+k_2\sin\theta_2\cos\phi_1\right)
E^{(1)}_{\mathbf{k}_1}E^{(1)}_{\mathbf{k}_2}.
\eqe
Here I used the coordinate system such that $k_y=0$ so that
$\mathbf{k}=k(\sin\theta,0,\cos\theta)$,
$\mathbf{k}_i=k_i(\sin\theta_i\cos\phi_i,\sin\theta_i\sin\phi_i,\cos\theta_i)$.
Transforming,
according to eq. (\ref{a}), the wave amplitudes from $E_\mathbf{k}$ to $a_\mathbf{k}$, yields the
expression for the matrix coefficient
\eqb
V^{\rm S\to A+A}_{\mathbf{k}_1\mathbf{k}_2\mathbf{k}}=
i\left(\frac{2\pi\omega_1\omega_2c}{k}\right)^{1/2}
\frac{k_1\sin\theta_1\cos\phi_2+k_2\sin\theta_2\cos\phi_1} {B_0}.
\eqe

One sees from the resonance conditions and the dispersion relations, that the frequencies of all
three interacting waves are generally of the same order and the angles are generally of the order
of unity. Then one gets a rough estimate for the matrix coefficients
\eqb
\vert V^{\rm S\to A+A}_{\mathbf{k}_1\mathbf{k}_2\mathbf{k}}\vert\sim \vert V^{\rm S\to
S+A}_{\mathbf{k}_1\mathbf{k}_2\mathbf{k}}\vert\sim \frac{\omega^{3/2}}{B_0}. \label{V}\eqe

\end{document}